\documentclass[aps,amsmath,amssymb,superscriptaddress,twocolumn,showpacs]{revtex4}
\usepackage{graphicx}
\usepackage{dcolumn}
\usepackage{bm}
\usepackage{colordvi}
\usepackage{color}

\begin{document}
\date{\today}
\title{Effect of gate-driven spin resonance on the conductance of a one-dimensional quantum wire}
\author{Almas F.~Sadreev}
\affiliation{L.V. Kirensky Institute of Physics, 660036,
Krasnoyarsk, Russia}
\author{E. Ya. Sherman}
\affiliation{Department  of  Physical Chemistry, Universidad  del
Pais Vasco UPV-EHU,  48080  Bilbao, Spain}
\affiliation{IKERBASQUE, Basque Foundation for Science, Bilbao, Spain}
\begin{abstract}
We consider quasiballistic electron transmission in a one-dimensional quantum wire
subject to both time-independent and periodic potentials of a finger
gate that results in a coordinate- and time-dependent Rashba-type spin-orbit coupling.
A spin dependent conductance is calculated as a function of external
constant magnetic field, the electric field frequency, and potential
strength. The results demonstrate the effect of the gate-driven electric dipole spin resonance in a
transport phenomenon such as spin-flip electron transmission.
\end{abstract}
 \maketitle
\section{Introduction}

Since the {Datta-Das spin} field-effect transistor \cite{datta} was
proposed, the Rashba spin-orbit interaction (RSOI) \cite{rashba}
has attracted considerable attention on account of its possible
applications in spintronics. The manipulation of electron spins can be
achieved via an external active control, which is the
essential requirement for spintronics devices. Interest in the RSOI as an
instrument to electrically manipulate spins in nanosystems \cite{Koo} has been growing since
Nitta {\it et al.} \cite{nitta} showed that in an inverted
In$_{0.53}$Ga$_{0.47}$As/In$_{0.52}$Al$_{0.48}$As quantum well the RSOI can  be
controlled  by  applying  a  gate  voltage. { In general, this control
is strongly material and structure-dependent, as was demonstrated
in more recent experiments on $n$-type semiconductors \cite{Heida,Engels,Shapers,Hu,Grundler,Eldridge}.
A similar effect of electric field has also been
achieved in a $p$-type InAs semiconductor, as reported by Matsuyama {\it et al.} \cite{Matsuyama}.}

\begin{figure}
\includegraphics[scale=.45]{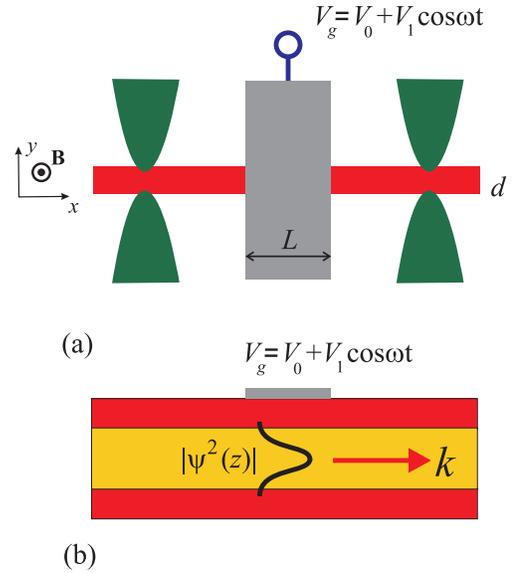}
\caption{(a) View of the { conducting channel} subject to a dc potential of the quantum
point contact gates and dc and ac potential of a finger gate. The
inhomogeneous electric field created by the finger gate (shown in
gray) represents the scattering active region. The distance between pairs of quantum
point contacts, $L_{w}$, is not shown here.
(b) Cross-section
of the nanostructure. The internal layer shows the propagation
channel, $k$ is the electron momentum, and $\psi(z)$ is the wave
function of the localized electron.
{ The gates for $y-$ axis confinement are not presented. The details of the figure
are not to scale.}
} \label{fig1}
\end{figure}

Assuming that a finger gate with dc voltage is located above a {
conducting channel based on a} two-dimensional electron gas (2DEG)
as shown in Fig. \ref{fig1} one can see that its electric field
gives rise to a local RSOI. Under the assumption of a stepwise RSOI the
electron ballistic transport in a quasi one-dimensional wire
has undergone a thorough investigation
\cite{Pershin,wang0,Zhang,Serra,sanchez,gong,sanchez1,Sris,Chirolli,Sigurdur,Yue}.
In the present paper we consider an actual electric field
produced by the dc biased finger gate that gives rise to a nonuniform RSOI. 
Next, we assume that the finger gate is also biased
by time-dependent (ac) voltage that affects the
space and time-dependent RSOI. Electron transport through
wires with spin-orbit interaction subjected to a time-periodic
potential was studied in Refs. \cite{Tang,Wu}.

An ac biased finger gate contributes to the time-periodic RSOI which
may give rise to many interesting effects such as generation of spin
current \cite{Malshukov,Wang,Tang1} and spin-polarized wavepackets \cite{Bogar}. 
A well-known and particularly powerful way of manipulating spins in doped III-V
heterostructures is electric dipole induced spin resonance (EDSR)
\cite{RE,Rashba,ER,Golovach06,Riu}  or gate-driven resonance
\cite{Laird} where the fields coherently driving the spins are
electric rather than magnetic such as in a conventional paramagnetic
resonance. Nowack  {\it et al.} \cite{Nowack} observed EDSR in a
single GaAs quantum dot and found that, as expected, the Rabi
frequency for spin flips is much less than the corresponding
Zeeman splitting. Kato {\it et al.} \cite{Nature} manipulated
electron spins in a parabolic AlGaAs quantum well by a GHz bias
applied to a single gate producing a field $E(t)$ perpendicular to
the well. Pioro-Ladriere {\it et al.} \cite{Pioro} studied the
effect of a slanting magnetic field for the EDSR. All these effects
were observed for electrons localized in quantum dots.
Although it is clear that the electron spatial dynamics, e.g., in
a double quantum dot can strongly modify the Rabi frequency of the EDSR \cite{Khomitsky,Novak}, 
there is no understanding of the signatures of the EDSR in the electron
transport. Here we study ballistic electron transport in
a one-dimensional (1D) quantum wire subject to dc and ac bias produced by a
finger gate and a Zeeman magnetic field and demonstrate the role of
the EDSR in the conductance of such a system. We found that the
effect leads to avoiding crossing in the dependence of the
conductance on the electron energy and ac field frequency, or
equivalently, the Zeeman splitting.


\section{Description of model}

The semiconductor structure of our interest is shown in Fig. \ref{fig1}
which demonstrates that { in the absence of a gate bias}
the system is symmetric  { with respect to the $z\to -z$ reflection}.
Then the system has only the Dresselhaus SOI because of the
host crystal electric field \cite{dressel,pikus},
$H_D=i\beta\langle
k_z^2\rangle[\sigma_x\partial_x-\sigma_y\partial_y],$ over the
whole sample. Next, similar to Ref. \cite{Tang2}, we consider a metallic
gate at a height $h$ from the { conducting channel} whose length
$L_{w}$ along the transport $x$-axis exceeds the gate width $L,$ { such that their
ratio $l\equiv L_{w}/L\ge 1$.} The
electrostatic potential of the biased metallic gate was derived by
Davies {\it et al.} \cite{davies}. For the structure presented in Fig. \ref{fig1}, the gate potential has
the form
\begin{eqnarray}\label{gatepot}
    V(x,z,t)=(V_0+V_1\cos\omega t)\phi(x,z),\nonumber\\
    \phi(x,z)=\frac{1}{\pi}\left[\arctan\frac{L+x}{z}+\arctan\frac{L-x}{z}\right],
\end{eqnarray}
where the gate is chosen as the origin of $z$. The gate potential 
produces the RSOI nonuniform over $x$, \cite{drell,winkler}
\begin{equation}
\label{HR} H_R=
-i{\alpha}\left[{\bf E}({\bm \sigma}\times{\bf {\bm\nabla}})-\frac{1}{2}{\bm\sigma}({\bm\nabla}\times {\bf E})\right],
\end{equation}
where ${\bf E}(x,y,z,t)=-{\bm\nabla} V(x,z,t)-{\bm\nabla} V_{\rm LC}(y)$. Here
$V_{\rm LC}(y)$ is the lateral confining potential \cite{moroz}, and the
last contribution in Eq. (\ref{HR}) ensures the hermiticity of the
RSOI. {The $y$-axis confinement length $d$ is typically tens
to hundreds of nanometers. This small length
establishes a corresponding high energy gap for the transverse excitations and
protects the system from exciting the transverse modes. In what follows we adopt
strong lateral confinement of about tens of nanometers in order to focus on
the effects of the ac potential for the $x$-axis electron transmission through
the one-dimensional quantum wire.
Then we can restrict ourselves to the ground state $\psi_0(y,z)$ which
is a sharp function compared to the characteristic scale along the wire $L_{w}$ which is taken to be 
of the order of hundreds of nanometers. }

The projection of the total Hamiltonian onto that ground state gives us the following effectively
one-dimensional Hamiltonian
\begin{eqnarray}\label{Hred}
   \widetilde{H}&=&\int dydz\psi_0(y,z)H\psi_0(y,z) \\
&=&\varepsilon_{0}
\left[\widetilde{H}_0+\widetilde{V}_0(x,t)+\widetilde{H}_Z+\widetilde{H}_R\right].\nonumber
\end{eqnarray}
Here
\begin{eqnarray}
&&\widetilde{H}_0=-\frac{\partial^2}{\partial x^2},\\
&&\widetilde{V}_0(x,t)=(v_0+v_1\cos\omega t)\phi(x,z=h)
\end{eqnarray}
are the dimensionless Hamiltonian of free motion of electrons and
the dimensionless potential of the finger gate, respectively.
The coordinate $x$ is measured in terms of the gate width $L$ and
the energy is measured in units of $\varepsilon_{0}={\hbar^2}/{2m^{*}L^2}$.
The Zeeman contribution is
\begin{equation}\label{Zeem}
 \widetilde{H}_Z=B\sigma_z
\end{equation}
where $B={g}\mu_{B}H_{\rm ext}/2\varepsilon_{0}$ is the
dimensionless magnetic field applied perpendicular to the wire as
shown in Fig. \ref{fig1} (a).  Here $g$ is the $g$-factor,
and $H_{\rm ext}$ is the magnetic field. We assume
that the magnetic length is much larger than the channel width $d$ and
neglect the influence of the magnetic field on the orbital motion.

\begin{figure}
\includegraphics[scale=.35]{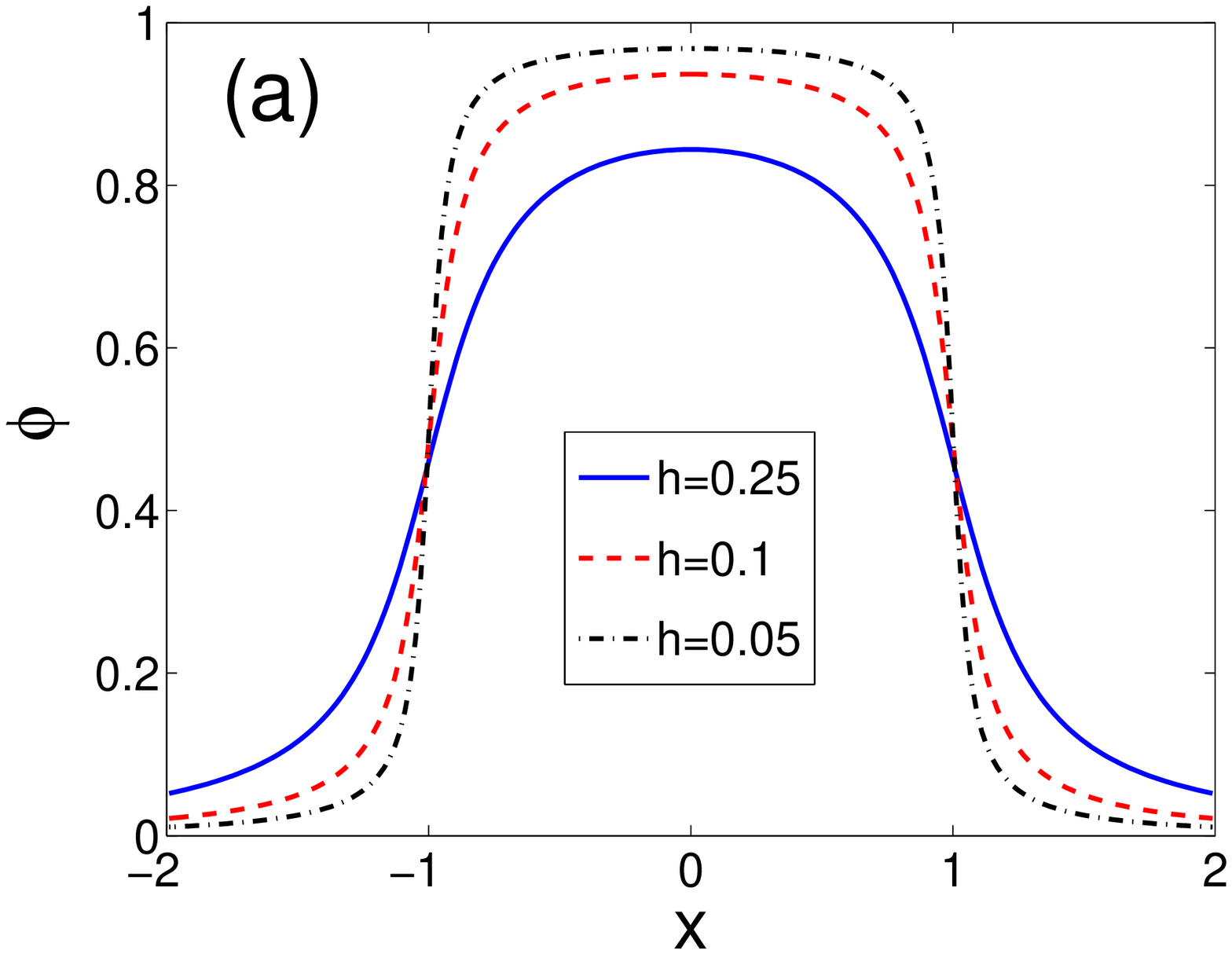}
\includegraphics[scale=.35]{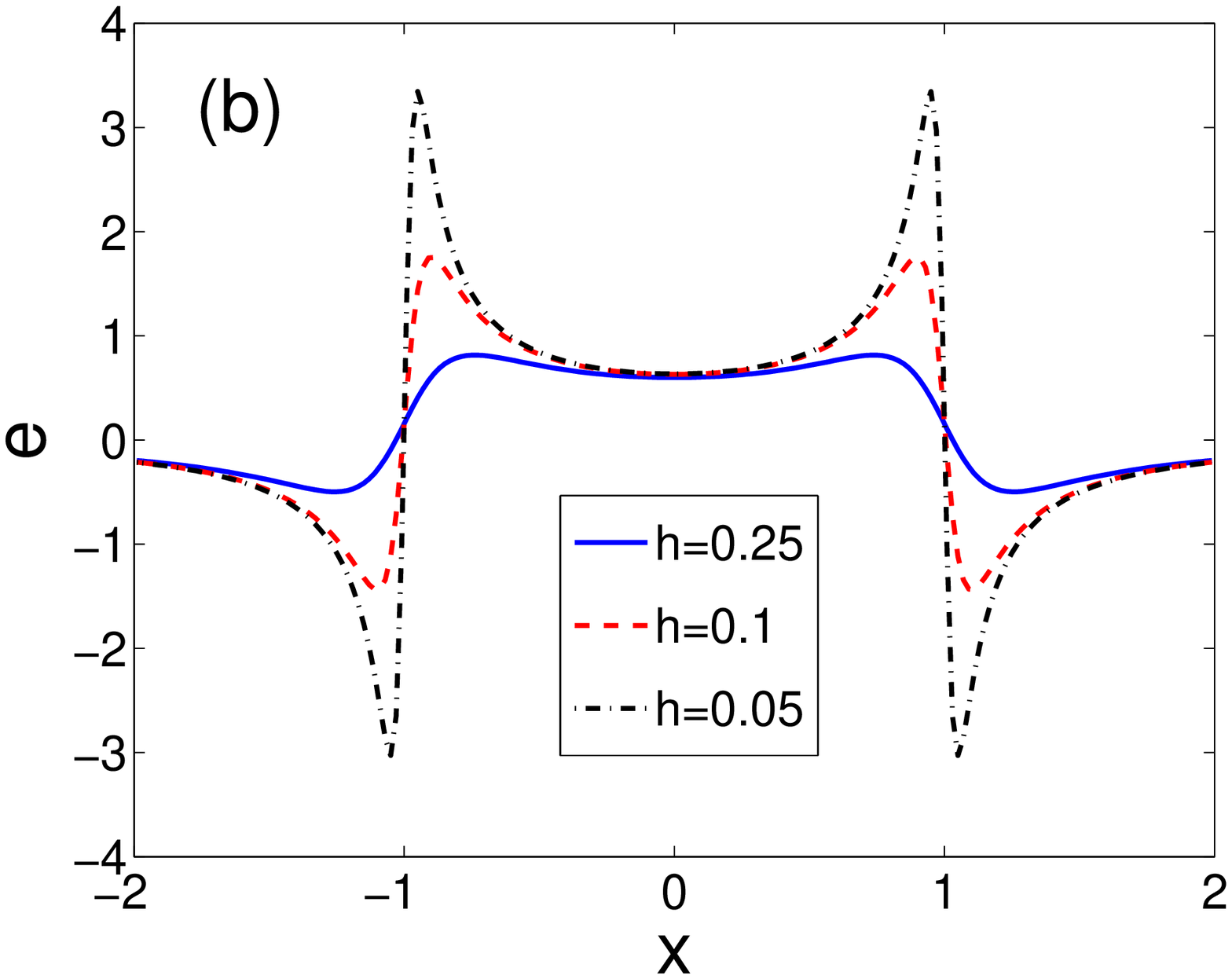}
\caption{Profiles of the dc potential $\phi(x)$ and electric field
$e(x)$ which define the non-uniform Rashba SOI in Eq.
(\ref{alpha}) over the transport axis for different distances
between the finger gate and the 2DEG. The potential is measured in
terms of $\varepsilon_{0}$ given here.} \label{fig2}
\end{figure}

The term
\begin{equation}\label{RSOI}
\widetilde{H}_R=i\varepsilon_{0}\sigma_{y}\left[\widetilde{\alpha}(x,t)\frac{\partial}{\partial
    x}+\frac{1}{2}\frac{\partial\widetilde{\alpha}(x,t)}{\partial
    x}\right]
    \end{equation}
constitutes the time-periodic RSOI where
\begin{eqnarray}\label{alpha}
&&\widetilde{\alpha}(x,t)=\widetilde{\alpha}(v_0+v_1\cos\omega t)e(x),\\
&&e(x)=\frac{1}{\pi}\left[\frac{1+x}{h^2+(1+x)^2}+\frac{1-x}{h^2+(1-x)^2}\right],
\end{eqnarray}
as follows from the dc and ac gate potentials (\ref{gatepot}). 
We use dimensionless variables with 
$\widetilde{\alpha}=(1\mbox {V}\times\alpha)/{\varepsilon_{0}L^2}$, and for practical
purposes take $1\mbox {V}/L$ as the unit of the electric field.
All distances are measured in terms of $L$, the magnetic field is
measured in terms of $2\varepsilon_{0}/g\mu_B$ and the frequency
$\omega$ is measured in terms of $\varepsilon_{0}/\hbar$,
respectively. We assume $\hbar\equiv1$ below in the text, if not explicitely
stated otherwise. Profiles of the
dimensionless electric field $e(x)$ are plotted in Fig. \ref{fig2}
for different distances $h$ of the finger gate from { the channel,} which shows
that the RSOI mainly contributes at the edges of the finger gate
for small height $h$. We employ here the representation:
\begin{equation}
\sigma_x=\left(\begin{array}{cc}
0 & -i\cr -i &0\end{array}\right),\quad \sigma_y=\left(\begin{array}{cc}
1 & 0\cr 0& -1\end{array}\right),\quad \sigma_z=\left(\begin{array}{cc}
0 & 1\cr 1&0\end{array}\right).
\end{equation}
The anticommutator form $({i}/{2})\left\{\widetilde{\alpha}(x,t),
{\partial}/{\partial x}\right\}$ in Eq. (\ref{RSOI}), which is related to
the inhomogeneous Rashba field $\widetilde{\alpha}(x,t)$ is often
adopted { by a phenomenological application of the Dirac
symmetrization rule for a product of noncommuting operators or it is 
taken for granted} \cite{lin,sanchez}. In addition, one can see that the RSOI caused by the
lateral confinement  is excluded because the electric field at the
position $y=0$ of the thin wire vanishes.

The values of the quantities necessary to describe the transport
are collected in Table I. To be specific,
we consider typical Rashba and Dresselhaus SOI
constants, effective masses, and $g$-factors for the InAs- and InSb-based heterostructures \cite{winkler,Silva}.
As seen from the table, the Dresselhaus SOI can be neglected in these
semiconductor structures even at rather weak applied fields.
In addition, we present the characteristic Zeeman field
and frequency corresponding to $\varepsilon_{0}$ for the finger gate length $L=1000$ \AA.

\begin{table*}
\caption{Parameter sets of the InAs- and InSb based
heterostructures for the gate length $L=1000$ \AA.} \label{tab1}
\begin{tabular}{|c|c|c|c|c|c|c|c|c|c|} \hline
 Structure &$m^{*} [m_0]$  & $\alpha$ [$e$\mbox{\AA}$^2$] &
 $\beta$ [eV$\cdot$\AA$^{3}$] &  $\varepsilon_{0}$ [meV]& ~~~~$\widetilde{\alpha}$~~~~
 &~~$\widetilde{\beta}$~~
 &~~~$g$~~~ & ~~~$B$ [for $H_{\rm ext}=1{\rm T}$]~~~ & $\omega/2\pi$ [GHz]\cr
  \hline
 InAs &0.023 & 117  & 27  & 0.15 & 0.7& $1.8\times 10^{-4}$ & 8 & 1.55 & 36\cr
 \hline
 InSb &0.035 & 523 & 760 & 0.23 & 3  & $3.3\times 10^{-3}$ & -10 & 1.26 & 55\cr
 \hline
\end{tabular}
\end{table*}

We begin with a stationary transmission for $v_1=0$ and $B=0$ and assume solely for this example
that the gate covers the entire channel, that is, $l=1$.
In this geometry we achieve the resonance
transmission when the electron energy matches the corresponding
eigenenergy of the gated wire channel. We assume that the Rashba coupling
is homogeneous over the wire subjected to the homogeneous potential $v_0$
according to  Eq. (\ref{gatepot}). As a result, the dependence of the 
resonances on the applied potential becomes, in general, parabolic 
due to the linear contribution of the Coulomb field and quadratic
in $v_{0}$ contribution of the spin-orbit coupling. 

\begin{figure}[b]
\centering
\includegraphics[height=6cm,width=7.5cm,clip=0.4]{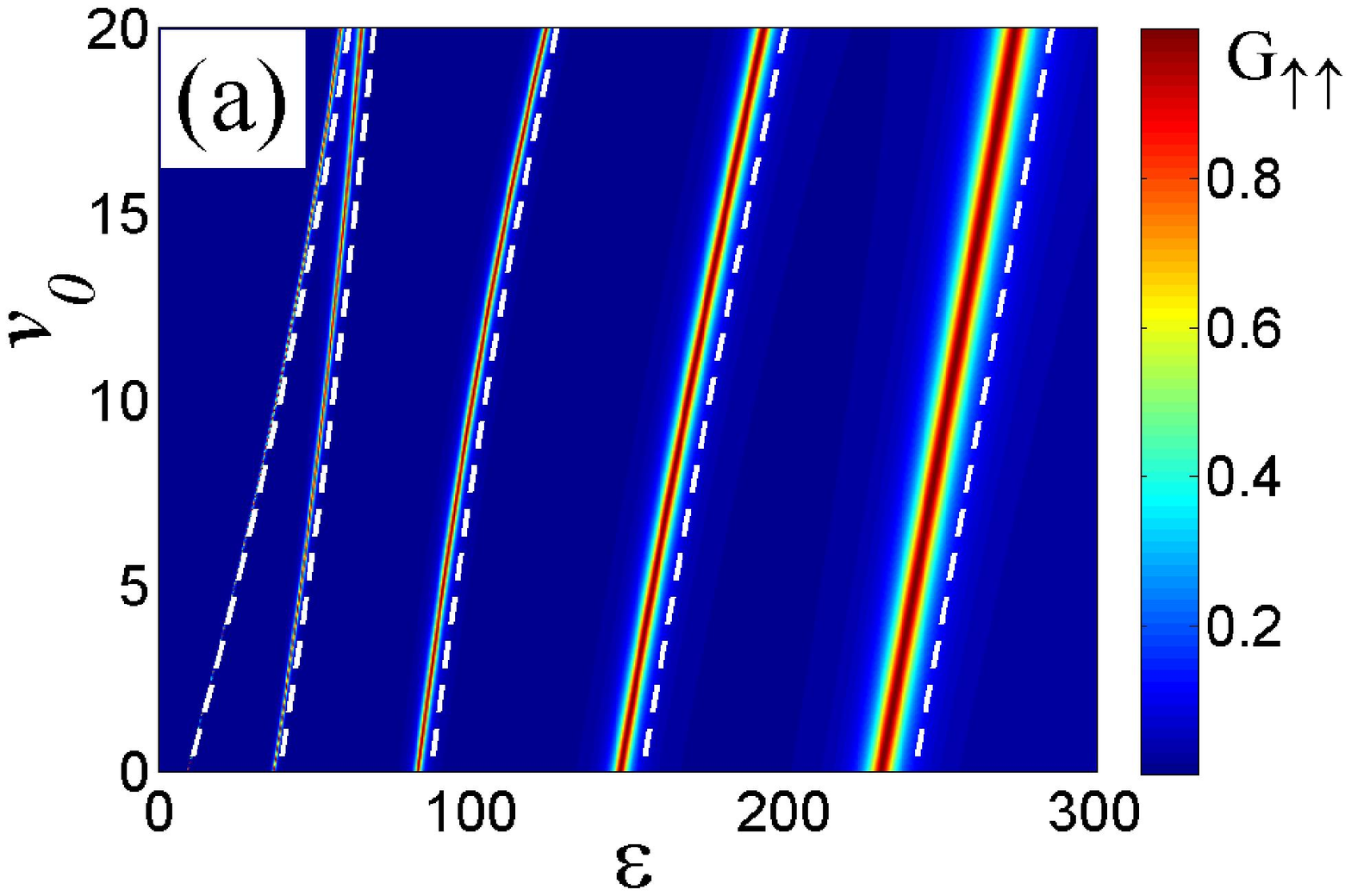}
\includegraphics[height=6cm,width=7.5cm,clip=0.4]{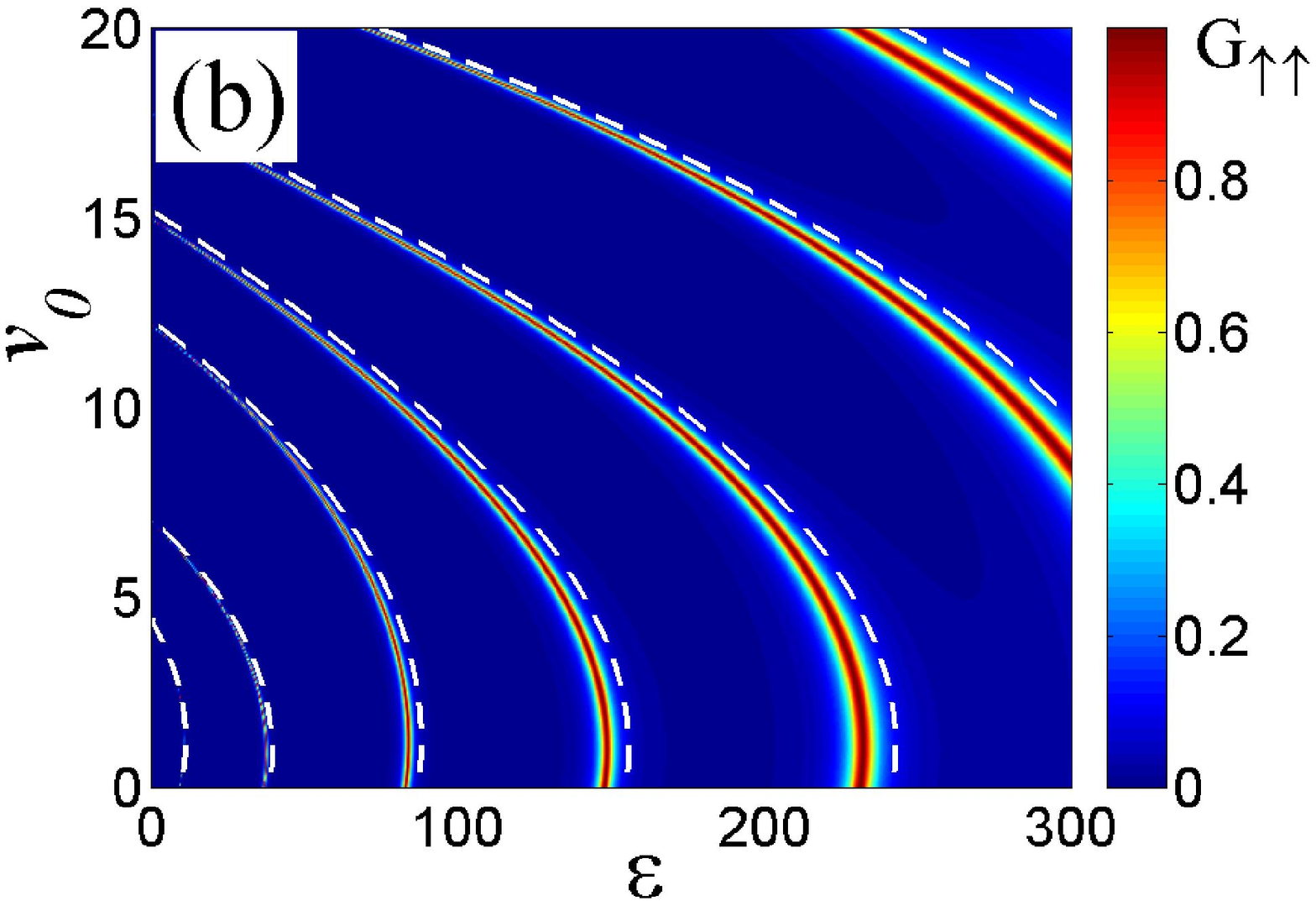}
\caption{Conductance $G_{\uparrow\uparrow}$ of the 1D wire vs
incident energy $E$ and dc potential applied to finger gate
(\ref{gatepot}) $v_0$ for $v_1=0, h=0.1, l=1$ (a)
$\widetilde{\alpha}=0$ and (b) $\widetilde{\alpha}=0.75$. The
incident energy and gate potential are measured in terms of
$\varepsilon_{0}$. Eigenenergies of closed 1D Rashba wire
for these parameters are shown by dashed white lines.} \label{fig3}
\end{figure}

We show the resonance peaks of the transmission which follow the
eigenenergies of the 1D Rashba box by dashed lines in Fig. \ref{fig3}. 
As expected, for $\widetilde{\alpha}=0$ the resonant transmission demonstrates 
linear behavior with $v_0$ [Fig. \ref{fig3}(a)].  For $\widetilde{\alpha}\neq 0$ 
the behavior of the eigenenergies of a closed wire with $v_0$ is parabolic [Fig.\ref{fig3}(b)]. 
Respectively, the resonance behavior of the Rashba wire demonstrates similar behavior as shown in Fig.
\ref{fig3} (b) for $\widetilde{\alpha}=0.75$.
In full agreement with the rigorous results in Refs. \cite{BS1,Xu}
the { numerically calculated} spin polarization
\begin{equation}\label{P}
    P=\frac{G_{\uparrow\uparrow}+G_{\uparrow\downarrow}-
    G_{\downarrow\uparrow}-G_{\downarrow\downarrow}}{G_{\uparrow\uparrow}+
    G_{\uparrow\downarrow}+G_{\downarrow\uparrow}+G_{\downarrow\downarrow}}
\end{equation}
vanishes because of the single-channel transmission in the 1D wire.

\section{AC assisted spin-dependent electron transmission}

Here we consider the spin-dependent transmission of electrons
through the 1D wire subjected to a dc and ac potential of the finger
gate (\ref{gatepot}). Before providing a detailed numerical
analysis, we address qualitatively an important question regarding whether a
spin polarization can appear in this situation in the absence of
an external magnetic field. For stationary single-channel
transmission in the quantum wire the RSOI cannot give rise to
spin polarization \cite{BS1,Xu}. However, the ac time-periodic
RSOI (\ref{alpha})  opens additional spin dependent channels of
electron transmission at the Floquet quasienergies
$\varepsilon+m\omega, m=0, \pm 1,\ldots$. Numerical calculations
in Refs. \cite{Wang,lin} show the spin polarization for the case
of the step-wise time-periodic RSOI. We argue that there is no
spin polarization for smooth coordinate behavior of the time-periodic
RSOI, at least, for zero Dresselhaus SOI. Indeed, for that case
the only spin component in the RSOI is $\sigma_y=\sigma,$ which is
an integral of motion. Then as Eq.(\ref{Hred}) shows the electron
transmission with spin $\sigma=1 (\uparrow)$ is not mixed with the
transmission with $\sigma=-1(\downarrow)$. From Eq. (\ref{Hred})
it follows that the electron transmissions with spin $\sigma=\pm1$ 
differ by only half of the time period $\pi/\omega$. And
therefore, after a time average of the conductances,
$G_{\sigma\sigma}$ do not depend on $\sigma$ while there is no 
conductance $G_{\sigma,-\sigma}$ with spin flip. Thus this
simple consideration proves that the spin polarization (\ref{P})
equals zero. Below we show that numerical computations agree with
that consideration for smooth space behavior of the time-periodic
RSOI.

The procedure of calculating the electron transmission through
the time-periodic potential (photon-assisted transmission) is well
described in literature \cite{landauer,jauho,wagner,BS,reichl,kohler}. 
There are two time-dependent contributions in our system. The first one 
is the periodic oscillations of the potential produced by the ac biased finger. 
This effect was considered in many publications for a spatially stepwise 
time-periodic potential \cite{landauer,jauho,wagner,reichl}. The
second one is the oscillating RSOI, 
considered for the stepwise spatial dependence of the ac finger 
field in Refs.[\onlinecite{Wang,lin}].

We use the tight-binding approximation to calculate the 
conductance through the space and time dependent profiles of the
potential \cite{kohler,sambe,peskin,sadreevPRE}. In the leads
where there is no SOI the wave functions can be written as
\cite{reichl,tien,valle,lefebvre},
\begin{eqnarray}
&&\mbox{left:} \nonumber\\
&&\psi_{j\sigma}(t) = \displaystyle{\sum_{m\sigma'}}\frac{e^{-i(\varepsilon+m\omega)t}}
{\sqrt{2\pi\rho(k_m)}}
[
\displaystyle{\delta_{m,0}\delta_{\sigma\sigma'}e^{ik_0j}+
\displaystyle{r_{m\sigma\sigma'}}e^{-ik_mj}}
], \nonumber\\
&&\mbox{right:} \nonumber\\
&&\label{leads}
\psi_{j\sigma}(t)=\displaystyle{\sum_{m\sigma'}}\frac{e^{-i(\varepsilon+m\omega)t}}{\sqrt{2\pi\rho(k_m)}}
\displaystyle{t_{m\sigma\sigma'}e^{ik_mj}},
\end{eqnarray}
where
\begin{equation}\label{kp}
\varepsilon+m\omega=-2\cos k_m,  ~\rho(k_m)=\partial \varepsilon/\partial k_m.
\end{equation}
Here we imply that the electron enters from the left lead with
energy $\varepsilon$ and spin $\sigma$ and reflects and transmits with
energy $\varepsilon+m\omega$ and spin state $\sigma'$ with corresponding
reflection and transmission amplitudes $r_{m,\sigma\sigma'}$ and $t_{m,\sigma\sigma'},$
respectively. We assume that the Zeeman and Rashba fields
affect the conducting electron in the 1D wire only. Inside the 1D wire of
length $L=a_0N$ with coordinate $x_j=a_{0}j, j=1, 2, \ldots, N$, we
present the Schr\"{o}dinger equation according to Eqs.
(\ref{Hred})-(\ref{alpha}) as follows:
\begin{widetext}
\begin{eqnarray}\label{SE2}
&&(\varepsilon+m\omega)\psi_{m,j\sigma}+\frac{\psi_{m,j+1\sigma}+\psi_{m,j-1\sigma}-2\psi_{m,j\sigma}}{a_0^2}
-v_{0}\phi_{j}\psi_{j,m\sigma}-v_{1}\phi_{j}(\psi_{j,m+1\sigma}+\psi_{j,m-1\sigma})-B\psi_{m,j,-\sigma}\nonumber\\
&& -i\widetilde{\alpha}\sigma
v_0e_j\frac{\psi_{m,j+1\sigma}-\psi_{m,j-1\sigma}}{2a_0}
-i\widetilde{\alpha}\sigma v_1
e_j\frac{\psi_{m+1,j+1\sigma}+\psi_{m-1,j+1\sigma}
-\psi_{m+1,j-1\sigma}-\psi_{m-1,j-1\sigma}}{2a_0}=0.
\end{eqnarray}
\end{widetext}
Here $\phi_{j}\equiv\phi(x_j,h)$ and $e_j\equiv e(x_j)$. To simulate entering and exiting through
quantum point contacts we implied the hopping matrix element $t=0.5$ between the leads and the
1D. In what follows we take the ratio $l=4.0$.

We define the transmission (reflection) probability as the ratio of
output current at the right lead to the input current, where
the current is defined in conventional units as
\begin{equation}\label{current}
J_{j\sigma\sigma'}=J_0\overline{{\rm
Im}[\psi^{*}_{j\sigma}\psi_{j+1,\sigma'}]}, ~~
J_0=\frac{e\hbar}{2m^{*}L}.
\end{equation}
Here
$\overline{\cdots}\equiv({\omega}/{2\pi})\int_0^{2\pi/\omega}\cdots
dt$. Substituting Eqs. (\ref{leads}) into Eq. (\ref{current}) we
obtain the dimensionless conductance
\begin{equation}\label{condtb}
G_{\sigma\sigma'}=\sum_m \frac{\sin {\rm
Re}(k_m)|t_{m\sigma\sigma'}|^2}{\sin k_0}.
\end{equation}
That expression reduces to the standard expression for the
conductance in the continuum approximation \cite{reichl}.
{ Taking the real part of $k_m$ in the nominator of Eq. (\ref{condtb})
assures that the Floquet states with  quasienergies $\varepsilon+m\omega$ beyond the
propagation band having an imaginary wave vector $k_m$ cannot participate in the conductance.}

\section{Numerical results}
In our numerical computations we chose the numerical lattice unit
$a_0=0.01$. For the dimensionless energies of electron $\varepsilon\sim 100$
the characteristic wavelength is of the order 1, which greatly exceeds
$a_0$. The next condition for $a_0$ is that $a_0\ll \Delta x$ where
$\Delta x$ is a characteristic scale over which the potential and
electric field of the finger gate undergo sharp changes as shown
in Fig. \ref{fig2}. This scale is close to the distance $h$
between the gate and { the channel}. Therefore the condition for the
numerical lattice unit is $a_0\ll h,$ which is satisfied also if we
take $h=0.1$. There is also the condition for the dimensionless
frequency $\omega>v_1/M$ where $2M+1$ is the number of Floquet
states \cite{BS,sadreevPRE}. In numerics we consider only the
minimal case $M=1$ to reduce the time of the numerical calculations
which means the region of small frequencies $\omega<v_1$ is excluded from 
consideration.
\subsection{Conductance in AC and DC biased wire}
\begin{figure}
\includegraphics[height=6cm,width=7cm,clip=]{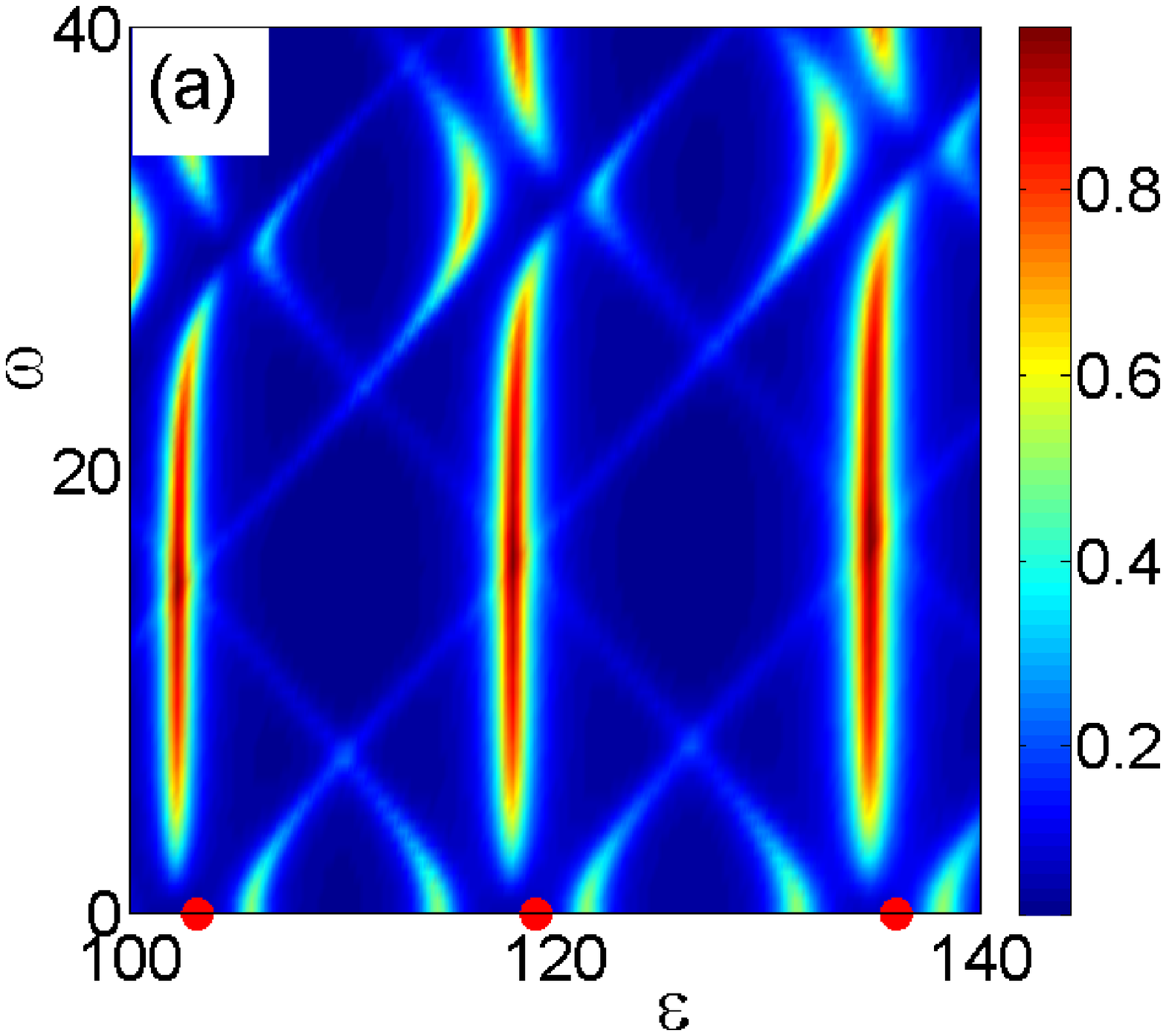}
\includegraphics[height=6cm,width=7cm,clip=]{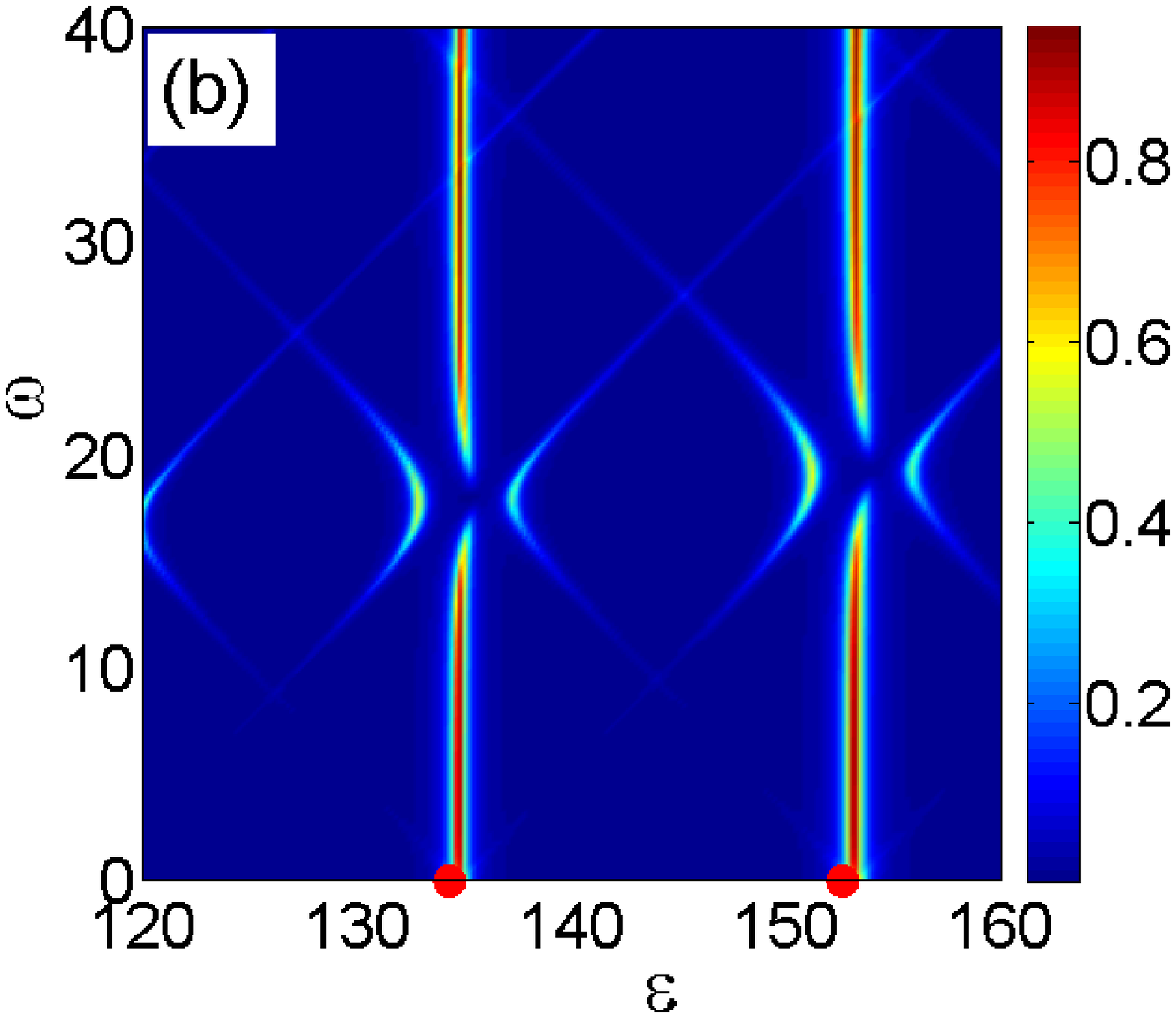}
\caption{Conductance $G_{\uparrow\uparrow}$ vs incident energy and
frequency of the ac potential for one gate with the parameters
$v_0=1, v_1=0.25, l=4$. (a) $\widetilde{\alpha}=0$ and (b)
$\widetilde{\alpha}=1$. The solid red circles mark the eigenenergies of the 
closed 1D wire.}\label{fig4}
\end{figure}

Figure \ref{fig4} shows the electron conductance of quantum wire
subjected to a finger gate that is dc and ac biased (Fig. \ref{fig1})
for the incident energies substantially exceeding the effective
potential barrier height. Switching off the ac potential,
$v_{1}=0,$ one obtains the resonance transmission for
$\varepsilon\approx \varepsilon_n$ where $\varepsilon_n$ are the
eigenenergies of the closed wire with an applied dc potential. These
eigenenergies are marked in Fig. \ref{fig4} by solid red
circles. The resonance positions, however, are slightly shifted because
of the openness of the 1D wire. Then the application of the ac potential
gives rise to the quasi energies
$\varepsilon_n\pm\sqrt{2v_1^2+\omega^{2}}$ \cite{sadreevPRE}.
The coincidence in these quasienergies with the basic eigenenergies
$\varepsilon_{n'}$ results in avoiding crossing as seen from Fig.
\ref{fig4} (a). Note because of the even symmetry of the potential
$\phi(x,z)$ relative to an inversion of the $x-$axis the only
Floquet states that avoid the basic energies are the ones which have the same
parity, i.e., $n'=n\pm2, n\pm 4, \ldots$ as seen from Fig.
\ref{fig4} (a).

Switching on the RSOI induced by the dc and ac electric fields
gives rise to the new selection rules of the avoiding crossing of
the Floquet states. The time-periodic term
$v_1\left\{\widetilde{\alpha}(x),k\right\}/2\times\cos\omega t$ in
the Rashba Hamiltonian (\ref{RSOI}) is odd with respect to the
$x$-inversion. Therefore it mixes the neighboring eigenstates
with the opposite parity of the closed wire while the time-periodic
potential $v_1\phi(x)\cos\omega t$ term mixes the eigenstates
with the same parity. As a result we obtain the avoiding crossings
of nearest neighbor resonances shown in Fig. \ref{fig4} (b). With the 
growth of the potential amplitude $v_0$ the avoiding crossings
occur irrespective of the selection rules that give rise to more
complicated frequency behaviors of the conductance.

It is possible to exclude the time-periodic perturbation of the 
potential by applying two oppositely biased finger gates symmetrically disposed
up and below the conducting layer. Then the electron experiences only
the time-periodic RSOI without a coordinate-dependent potential energy. 
For a given ac finger bias, here the effects of avoiding crossings are expected to be 
stronger due to the doubling of the electric field
affecting the RSOI.

\subsection{AC affected spin resonance for transmission in magnetic
field}
\begin{figure}
\includegraphics[height=6cm,width=7cm,clip=]{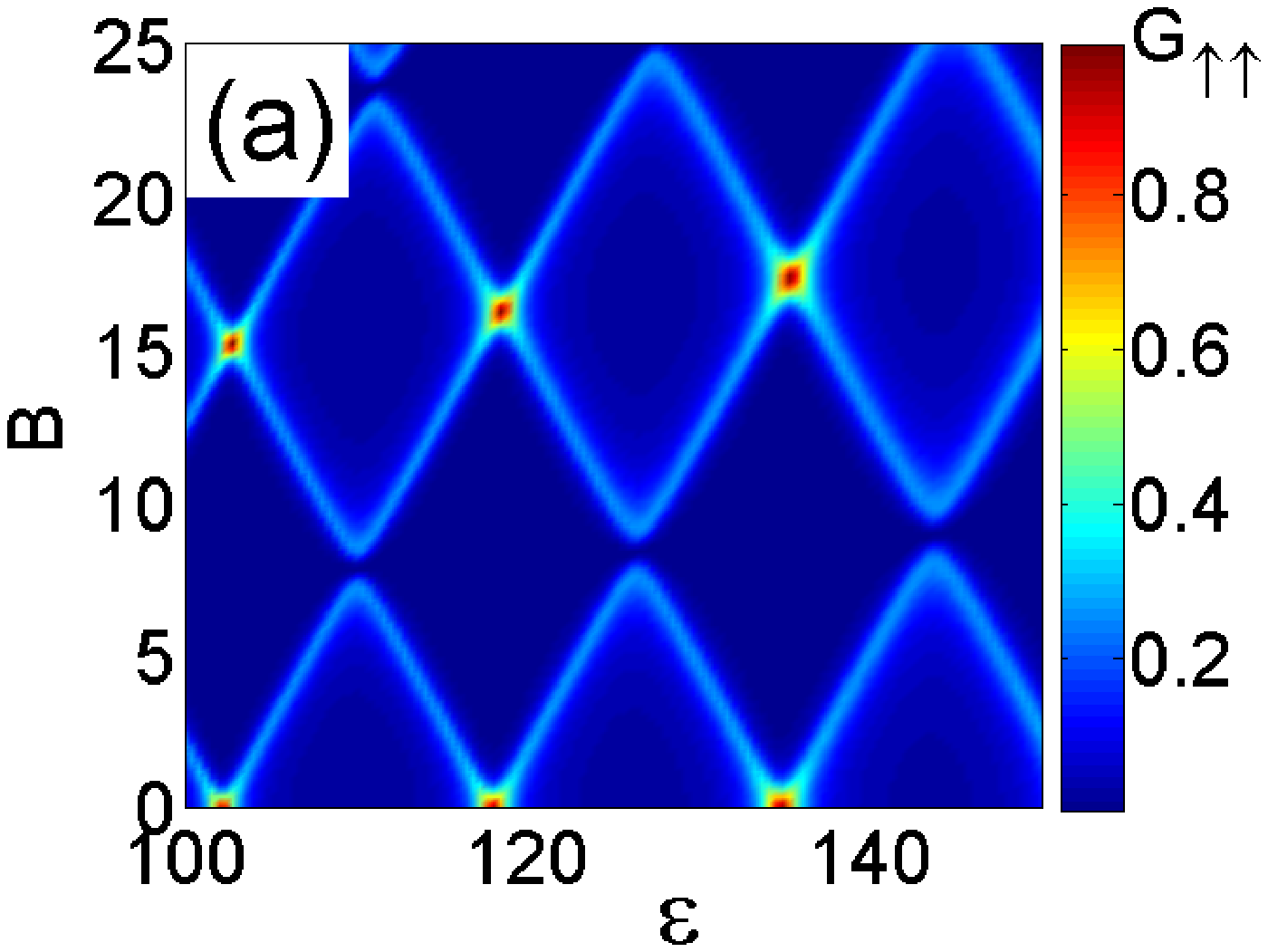}
\includegraphics[height=6cm,width=7cm,clip=]{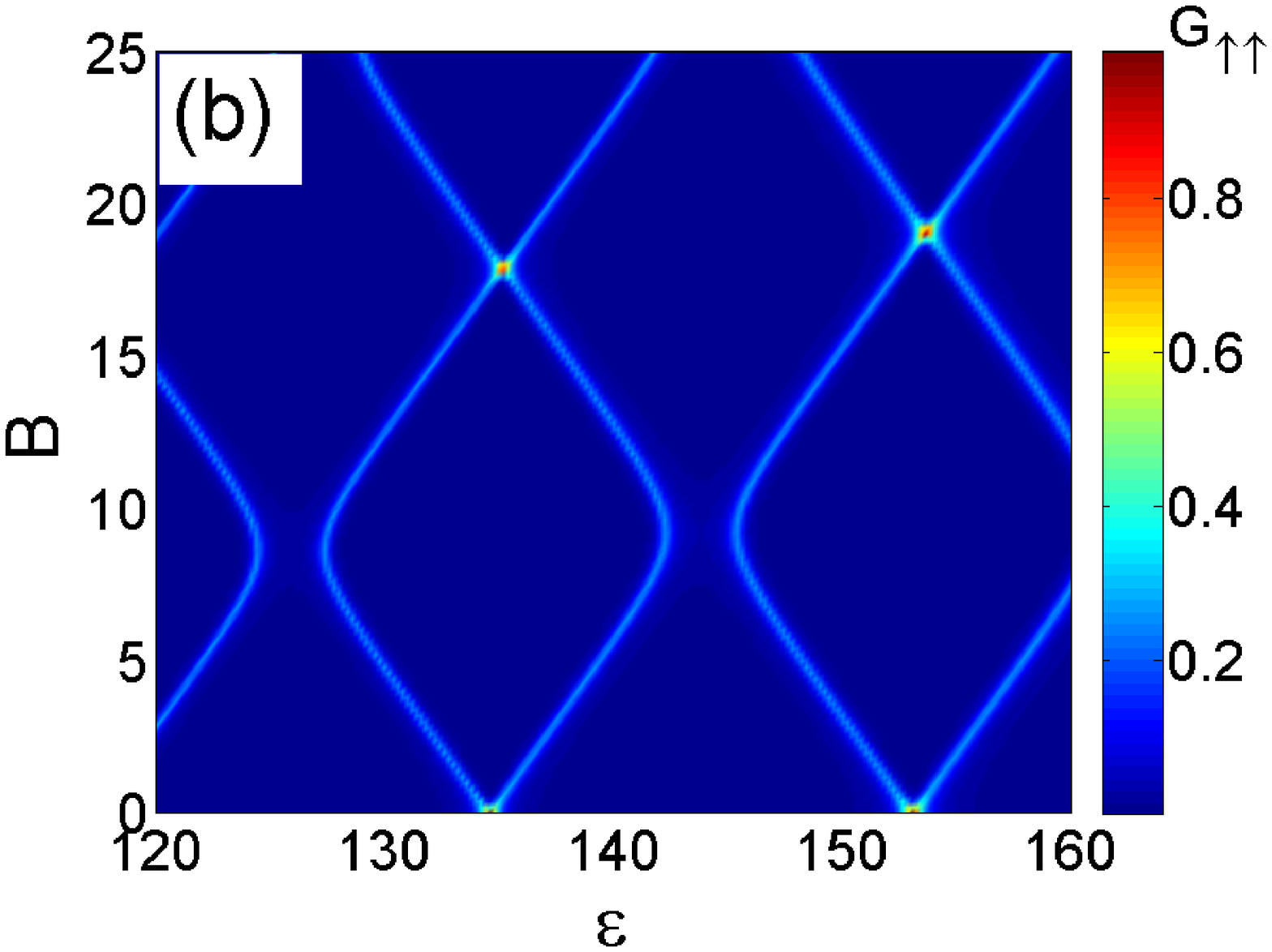}
\includegraphics[height=6.5cm,width=7.5cm,clip=]{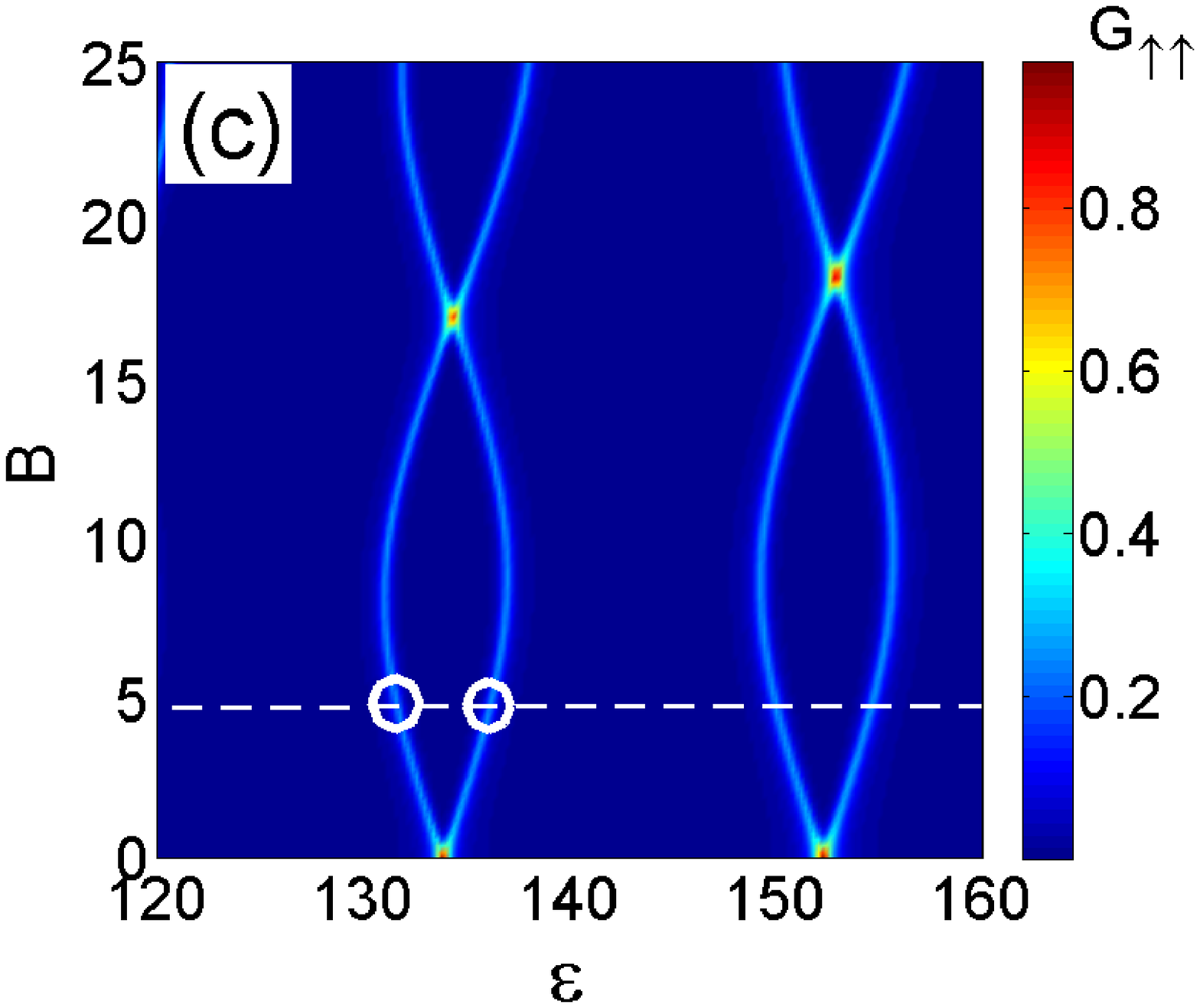}
\includegraphics[height=6cm,width=7cm,clip=]{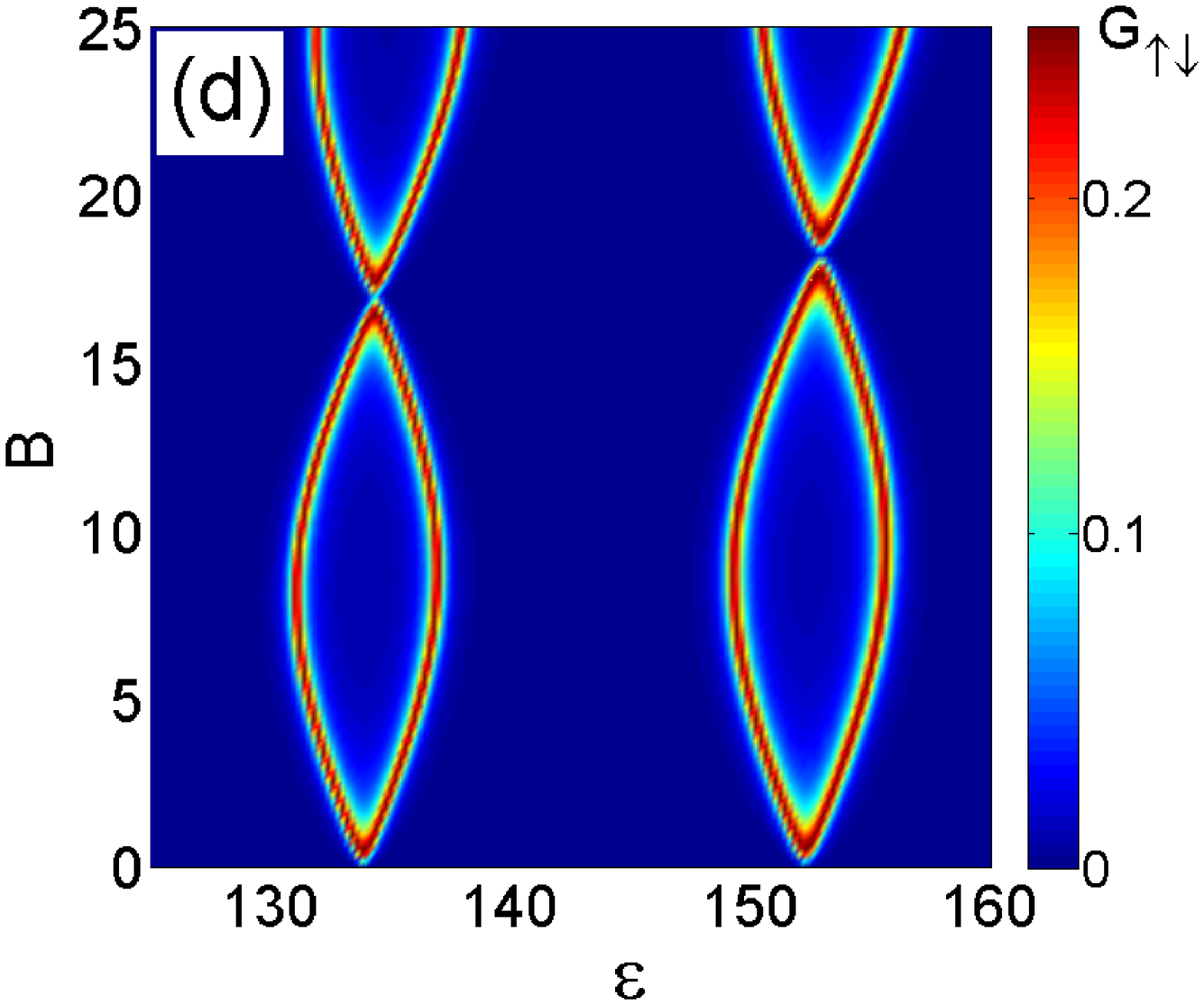}
\caption{The stationary spin-dependent conductances
$G_{\sigma\sigma'}$ vs magnetic field and energy for
$\widetilde{\alpha}= 0$ (a),  $\widetilde{\alpha}=0.25$ (b),
$\widetilde{\alpha}=1$ (c) and (d). $v_0=1$.} \label{fig5}
\end{figure}

Now we apply the magnetic field perpendicular to the quantum wire
as shown in Fig. \ref{fig1} (a). For the dc potential the
term (\ref{Zeem}) obviously gives rise to the Zeeman splitting of
the energy levels of the wire. Respectively, the resonance
transmission follows these split energy levels as shown in Fig.
\ref{fig5}. For $\alpha=0$ the conductance simply follows the magnetic
field as seen from Fig. \ref{fig5} (a), while the RSOI leads to
avoiding crossing behavior of the conductance because of
$[\widetilde{H}_R, \widetilde{H}_Z]\neq 0,$ as seen from Figs. \ref{fig5}
(b), \ref{fig5} (c), and \ref{fig5} (d).

The most important point is that the last term in the Hamiltonian
(\ref{Hred}) has similar effects as the radio frequency magnetic field
directed perpendicular to the constant Zeeman field. Therefore we can
expect signatures of spin resonance  for $\omega \approx 2B$
with spin inversion \cite{RE,Rashba,ER}.
\begin{figure}
\includegraphics[height=5.5cm,width=7cm,clip=]{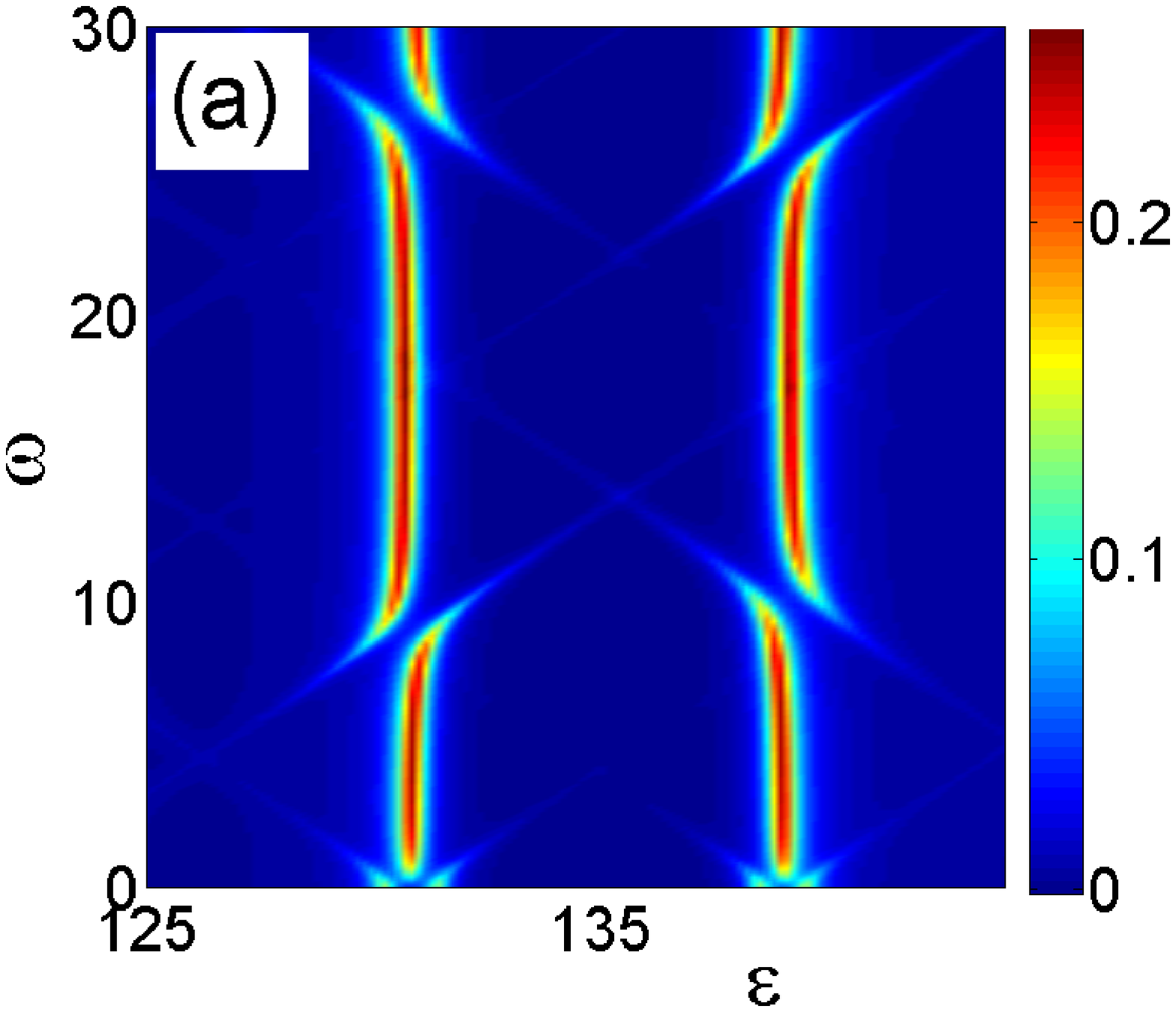}
\includegraphics[height=5.75cm,width=7cm,clip=]{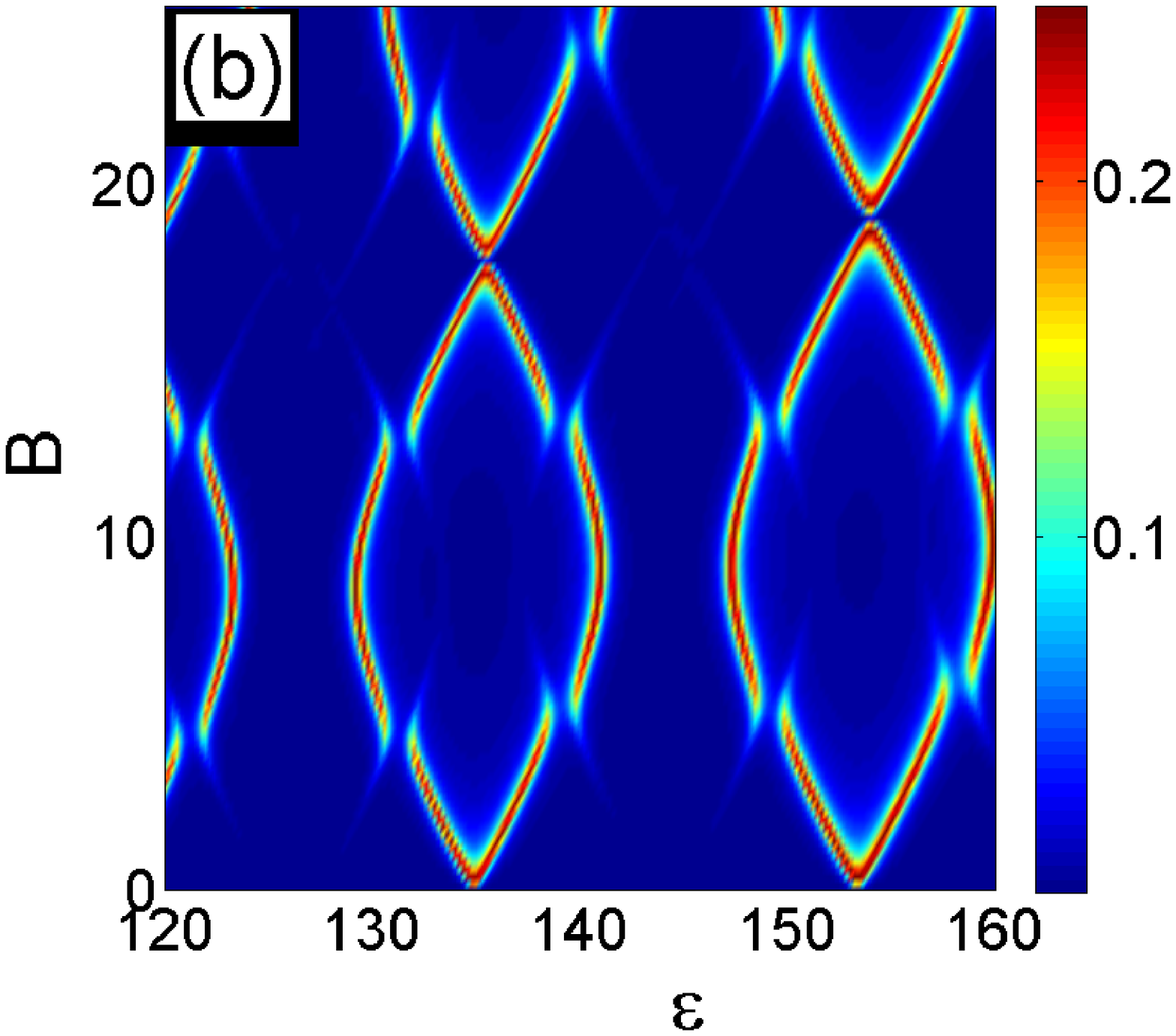}
\caption{The ac affected conductance $G_{\uparrow\downarrow}$ (a)
vs incident energy and frequency of the ac potential when
the magnetic field $B=5$ is applied perpendicular to the wire and (b) vs
incident energy and external magnetic field for $\omega=10$. The
parameters are $v_0=1, v_1=0.25, l=4, \widetilde{\alpha}=1$.
} \label{fig6}
\end{figure}
We take $B=5,$ which is shown by the dashed line in Fig. \ref{fig5} (c).
Figure \ref{fig6} (a) shows the conductance $G_{\uparrow\downarrow}$
vs energy and frequency of the ac potential with the RSOI
$\widetilde{\alpha}=1$. For $\widetilde{\alpha}=0$ this conductance is
zero and therefore is not presented.
One can see that the basic resonances in conductance follow the RSOI
and Zeeman split eigenenergies, shown in Fig. \ref{fig5} (c) as
open circles. However, there is a fine structure of the conductance
in the form of avoiding crossings where the Floquet resonances
cross the basic resonances, which are marked in Fig. \ref{fig5} (c) by the open
circles. That indicates spin resonances for swiping the
frequency of the ac potential.  Figure \ref{fig6} (b) shows the
conductance $G_{\uparrow\downarrow}$ for the fixed frequency of the ac
potential $\omega=10$ vs incident energy and a constant
magnetic field applied perpendicular to the transport axis $x$.
Similar to the case in Fig. \ref{fig6} (a) we see the self-avoiding
Floquet resonances with the basic Zeeman peaks of the
conductance shown in Fig. \ref{fig5} (d). This result
clearly shows spin resonances affected by the electric ac potential.

\begin{figure}
\includegraphics[height=7cm,width=7.5cm,clip=]{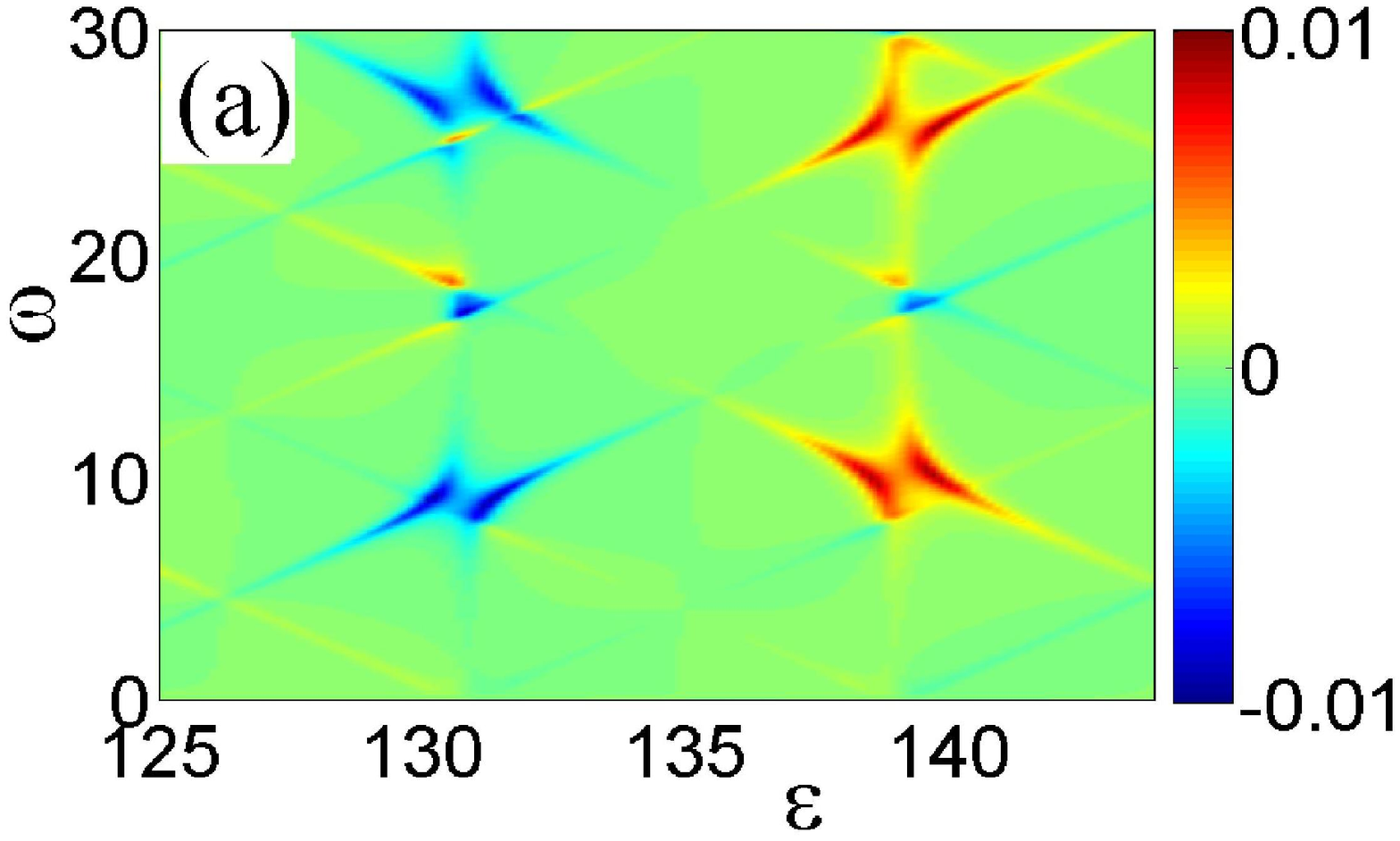}
\includegraphics[height=6.5cm,width=7.5cm,clip=]{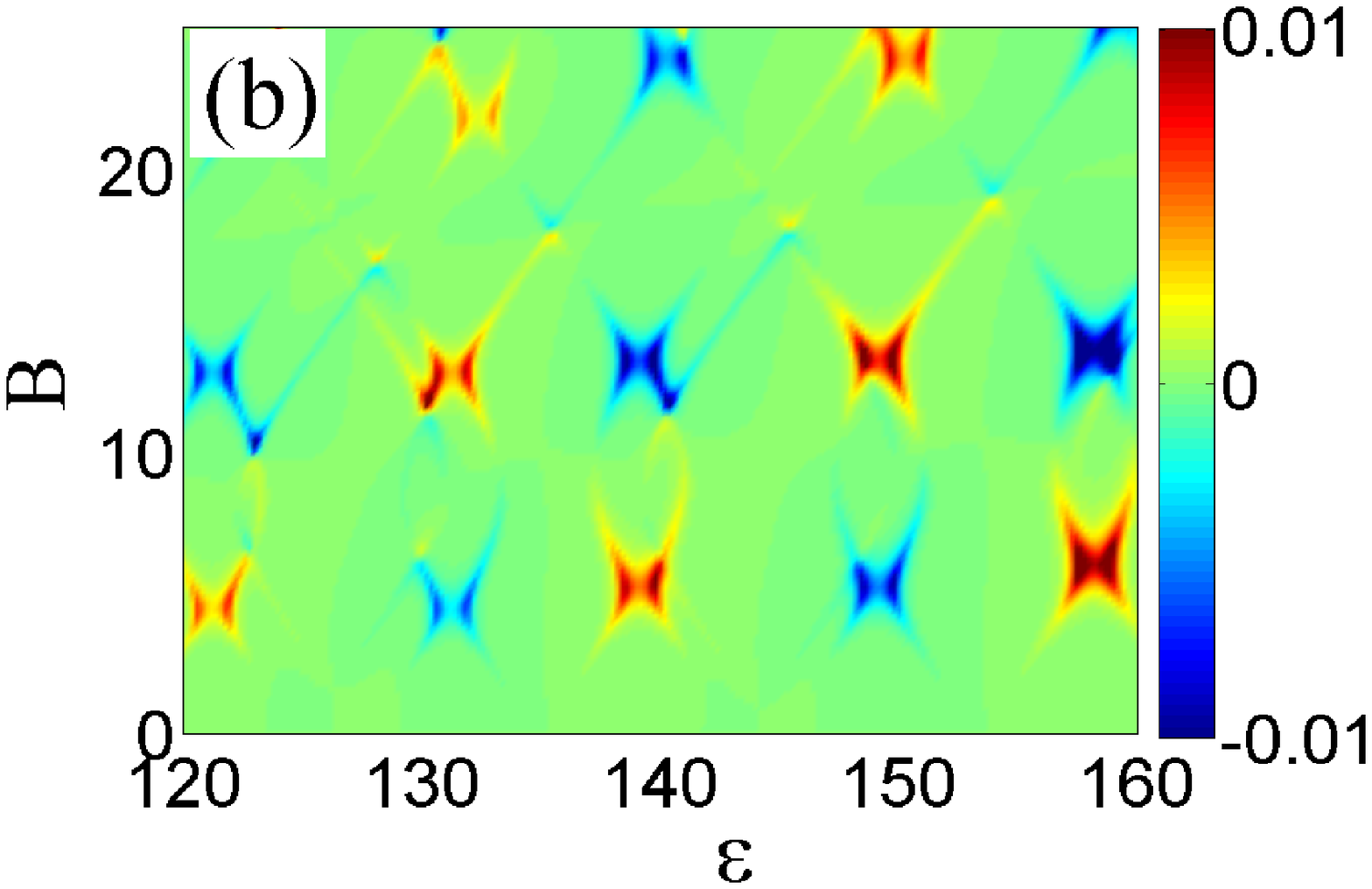}
\caption{The difference $G_{\uparrow\downarrow}$-$G_{\downarrow\uparrow}$
vs incident energy and frequency of the ac potential when the 
magnetic field $B=5$ is applied perpendicular to the wire and (b) vs
incident energy and an external magnetic field for $\omega=10$.}
\label{fig7}
\end{figure}

An interesting feature of the transmission in a nonzero magnetic field, where
the time-reversal symmetry is broken, is the difference between two spin-flip channels,
that is $G_{\uparrow\downarrow}-G_{\downarrow\uparrow}$. It is presented in Fig. \ref{fig7}
and corresponds to $G_{\uparrow\downarrow}$ in Fig. \ref{fig6}. Although this difference is
small, being of the order of 0.1 of $G_{\uparrow\downarrow},$
and appears mainly in the anticrossing spin-flip domains of Fig. \ref{fig6},
its nonzero value is the qualitative manifestation of the broken time-reversal symmetry,
and, as a result, of the possible generation of finite spin  polarization [see Eq.(\ref{P})].

\section{Summary and discussion}

We studied the effects of dc and ac biased finger gate on the
resonant transmission of an electron through a 1D quantum wire.
The potential and the electric field of the gate are local as
shown in Fig. \ref{fig2}. The ac field of the gate forms a
time-periodic Rashba SOI which can cause spin flip for the
electron that is transmitted through the gated channel. This results in
new features in the spin-flip electron conductance
$G_{\uparrow\downarrow}$ such as the Floquet satellites and
self-avoiding crossing of resonances, while the basic resonances
follow the eigenenergies of a closed 1D wire subject to a dc
potential, a Zeeman magnetic field applied across the wire, and
a static RSOI. The simplest resonance-induced transition
corresponds to the matching of the frequency-dependent Floquet
resonance peak with the basic resonance peak corresponding to the
Zeeman splitting. That results in a spin resonance that is similar to
that formulated by Rashba and Efros when the time-periodic
electric field gives rise to a spin flip in a constant magnetic
field with spatially uniform spin-orbit coupling \cite{ER}. As can
be seen in Fig. \ref{fig6}, the characteristic static dimensionless
magnetic field and gate frequency in our model consideration are
of the order of 10. According to Table \ref{tab1} these
numbers correspond to ten Tesla and a few hundreds GHz (in the
upper range of the microwave radiation), respectively. In
addition, the characteristic required electric fields are of the
order of $10^{4}$ V/cm. In practice, these parameters are strongly
system-dependent and the effect, studied here only semiquantitatively, 
can be possibly observed at lower fields and
frequencies.

As additional conclusions, we would like to comment on the possibility of producing  spin
polarization in electron single channel transmission by the ac field.
As it was argued in Sec. III
the time-periodic Rashba SOI cannot lead to spin polarization
for the transmission through a 1D wire at $B=0$. This result agrees
with our computer simulations but disagrees with the numerical
results of Ref. \cite{lin} where spin polarization around
0.2 was found for zero Dresselhaus SOI $\beta=0$. An
origin of the difference is related to the coordinate dependence of the
finger gate field. Numerical calculations show a tendency for
decreasing the spin polarization with decreasing the simulation lattice
constant $a_0$ and increasing the height $h$, when the electric
field and potential become smooth. At $a_0\ll h$ the spin
polarization becomes negligibly small. Thus the stepwise approximation of such a non uniform
RSOI conceals a danger for numerical computations based on
finite difference schemes.

The manifestation of the electric dipole spin resonance in the
ballistic transport through a one-dimensional channel can help in the 
design of devices with a spin transport controlled by an electric field
in quantum nanoscale and mesoscopic systems.

\section*{Acknowledgments}
We have benefited from discussions with E.A. de Andrada e Silva and
Roland Winkler. The work of AS was partially supported by the
RFBR grant 13-02-00497. EYS acknowledges support
from the University of Basque Country UPV/EHU under
program UFI 11/55, Spanish MEC (FIS2012-36673-C03-01), and ''Grupos
Consolidados UPV/EHU del Gobierno Vasco'' (IT-472-10).


\end{document}